\documentclass{aa}
\usepackage{graphics}

\usepackage{natbib}
\usepackage{graphicx}

\input{psfig}

\begin{document}

\title{Detection of Saturnian X-ray emission with XMM-Newton}

\author{J.-U. Ness, J.H.M.M. Schmitt, and J. Robrade}
\institute{
Hamburger Sternwarte, Universit\" at Hamburg, Gojenbergsweg 112,
D-21029 Hamburg, Germany}

\authorrunning{Ness et al.}
\titlerunning{XMM observation of Saturn}
\offprints{J.-U. Ness}
\mail{jness@hs.uni-hamburg.de}
\date{Received \today;accepted ...}

\abstract{The giant planet
Saturn was observed by XMM-Newton in September 2002. We present and analyse
these XMM-Newton observations and compare our findings to the {\it Chandra}
observations of Saturn. Contamination of the XMM-Newton data by optical light
is found to be severe in the medium and thin filters, but with the thick filter
all optical light is sufficiently blocked and the signal observed in the this
filter is interpreted as genuine X-ray emission, which is found to
qualitatively and quantitatively resemble Saturn's {\it Chandra} spectrum very
well.
}
\maketitle

\section{Introduction}
\label{intro}

Most of the larger solar system objects are now known to emit X-rays via some
variety of different X-ray emission mechanisms. The most prominent example is
the gas giant Jupiter, whose X-ray emission is dominated by auroral emission,
produced by charged particles entering the planet's magnetosphere
\citep[e.g.,][]{metzger83}. A recent {\it Chandra} observation of the gas giant
Saturn \citep{ness_sat}
resulted in a definitive detection of X-ray emission also from this planet,
thus confirming a tentative ROSAT detection reported earlier by \cite{ness00},
however, at a level much lower than observed from Jupiter. Further, unlike
Jupiter, Saturn's X-ray emission is not concentrated in the polar regions, and
in fact, the detected level of Saturnian X-ray emission is consistent with the
observed level of Jupiter's equatorial emission \citep{ness_sat,waite96}. The
emission mechanism consistent with spectral and spatial
properties of the observed X-ray emission was found to be elastic scattering and
fluorescent scattering of solar X-rays, however, for this to be the case,
the X-ray albedo of Saturn has to be unusually high. \cite{ness_sat} estimated
the X-ray albedo required to explain the measured X-ray flux by scattering
processes and found a value of $>5.7\cdot10^{-4}$, which is about a factor
50 higher than for the moon \citep{schmitt91}. Since models of combinations
of scattering processes for Jupiter's equatorial emission \citep{maur00}
underestimate the observed flux level \citep{waite97} by a factor of 10, the
scatter process scenario has to explain a high X-ray albedo in both cases,
Jupiter's equatorial emission and Saturn's total emission.

Saturn was also observed by XMM-Newton in September 2002. We present and analyse
these XMM-Newton observations and compare our findings to the {\it Chandra}
observations of Saturn. We concentrate on the detection and spectral properties
of the X-ray photons, while the light curve provides only little information
due to its short duration.

\section{Observations and Data Analysis}
\label{anal}

\subsection{Observations}

Saturn was observed with XMM-Newton on September 10, 2002 for a total of
60\,ksec. The observations were split in three separate parts of almost equal
length, with different filter settings used for the PN and MOS detectors.
The MOS detectors were operated with medium filter + full frame, medium
filter + large window, and thin filter + large window. At the position of
Saturn no X-ray photons can be extracted from the MOS detectors, because
pixels with high optical load are not read out. The RGS is not affected by
optical light contamination, but the count rate is too low to obtain useful
spectra. We therefore use only the EPIC-PN detector for our analysis, and the
observation details are summarized in Table~\ref{tab1}. We inspected all three
PN-observations, but the data taken with the thin filter are close to useless
and the data taken with the medium filter are severely contaminated by optical
light. However, in the data taken with the thick filter no obvious signs of
optical contamination are apparent, consistent with our expectations about the
optical blocking power of the thick filter.

\begin{figure}[!ht]
  \resizebox{\hsize}{!}{\includegraphics{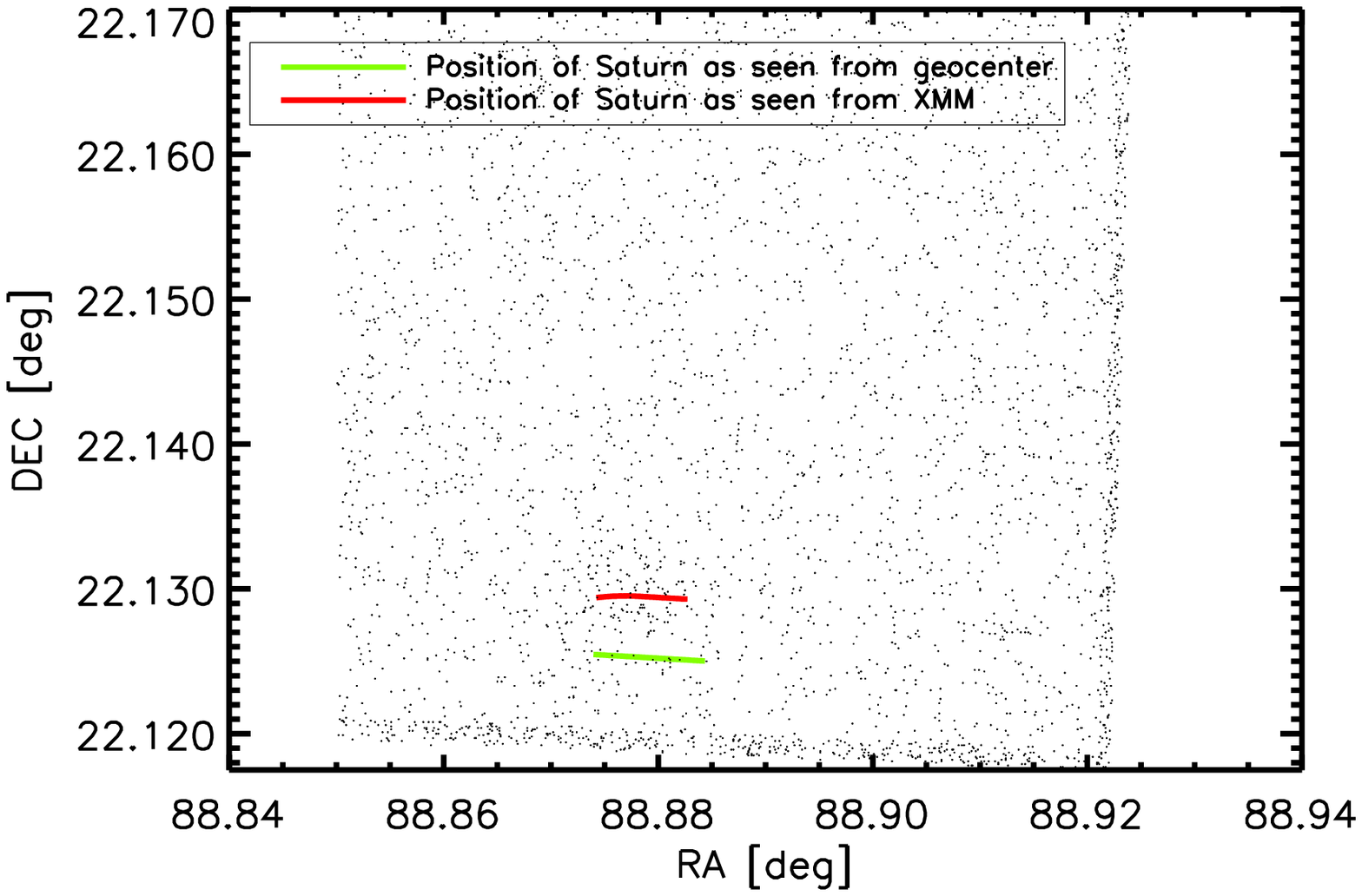}}

  \vspace{-1cm}\resizebox{\hsize}{!}{\includegraphics{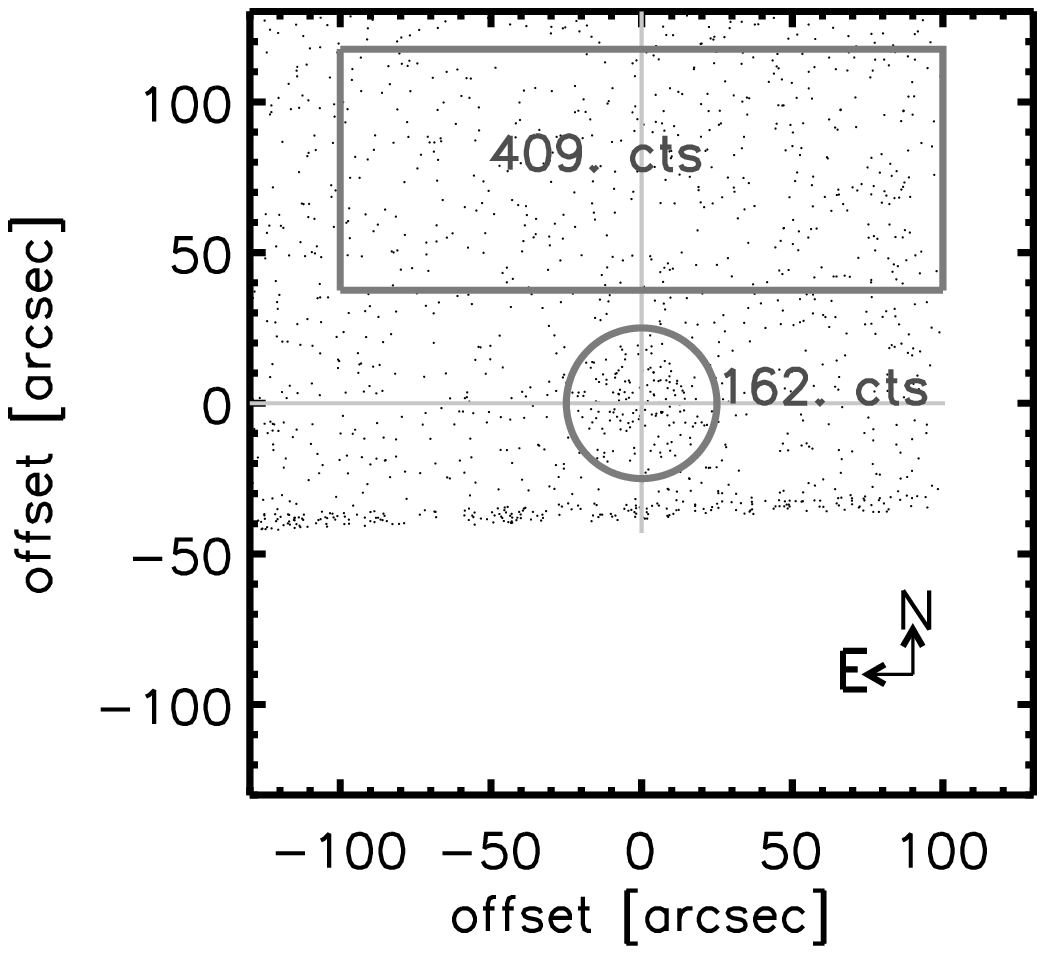}}
\caption{\label{pos}Photon positions on PN chip \#4 (upper panel; with expected
path of Saturn) and transformed positions in ``saturnocentric" coordinates
(bottom). 162 counts are extracted in a circle with radius 25\arcsec\ and 409
counts in an 80\arcsec$\times$200\arcsec\ background box.}
\end{figure}

Due to the small apparent motion of Saturn ($\sim 5$\arcsec, compare to the
instrument half-power diameter of 15\arcsec) during the 20\,ksec observation
interval and the high sensitivity of the PN-detector we could directly
identify Saturn on chip \#4 without any need of a transformation following
the apparent planetary motion (see Fig.~\ref{pos}, upper panel).
After having found emission at Saturn's position we carried out a transformation
procedure \citep[described in detail by ][]{ness_sat}, which transforms all
recorded events into a Saturnocentric coordinate system, and constructed an
image in this new coordinate system. We then extracted all photons within an
extraction radius of 25\arcsec\ around the central position of the transformed
coordinate system, where an enhancement of photons can be immediately
recognized (cf. Fig.~\ref{pos}). In a circular detect cell we extract
162 photons while from the background (extracted from an adjacent 80\arcsec
$\times$\,200\arcsec\ box) we expect only 50.2 photons. With
a total of 112\,$\pm$\,13 counts we have therefore obtained a highly
significant detection.

\begin{table}[!ht]
 \caption{\label{tab1}Overvation details for Saturn (only EPIC/PN).}
  \begin{tabular}{lr}
 \multicolumn{2}{l}{\bf ObsID 0089370501}\\
   Exp. time & 24024\,ksec\\
   Start Time & 2002-10-01 10:35\\
   Stop Time & 2002-10-01 17:15\\
   PN filter & THICK FILTER\\
   on-time (PN) & 21047\,ksec\\
 \multicolumn{2}{l}{\bf ObsID 0089370601}\\
   Exp. time & 24023\,ksec\\
   Start Time & 2002-10-01 17:35\\
   Stop Time & 2002-10-02 00:15\\
   PN filter & MEDIUM FILTER\\
   on-time (PN) & 20966\,ksec\\
 \multicolumn{2}{l}{\bf ObsID 0089370701}\\
   Exp. time & 24023\,ksec\\
   Start Time & 2002-10-02 00:37\\
   Stop Time & 2002-10-02 07:17\\
   PN filter & THIN FILTER1\\
   on-time (PN) & 20962\,ksec\\
\hline
   Angular diam. & 18.1\arcsec\\
   distance (Earth) & 9.2\,AU\\
   distance (Sun) & 9.0\,AU\\
   inclination  & -26.4$^\circ$\\
\hline
  \end{tabular}
\end{table}

\section{Results}

We analyzed the XMM-Newton EPIC-PN data of Saturn in the same fashion as the
{\it Chandra} data as described by \cite{ness_sat}. Since the recorded count
number is exactly the same (!), the detection significance in both data sets
is very similar; note, that the background values differ somewhat. Since the
angular resolution of the XMM-Newton data is lower, we are not able to locate
the X-ray emission on Saturn's apparent disk (diameter 18.1\arcsec) from our
XMM-Newton observations.\\
We extracted the light curve with different time bins but found no significant
variability. The net count rate is ($5.3\,\pm\,0.6)\cdot10^{-3}$\,cps. Only
half a rotation is covered with the short observation time and no phase
variability can be tested.

\begin{figure}[!ht]
\resizebox{\hsize}{!}{\includegraphics{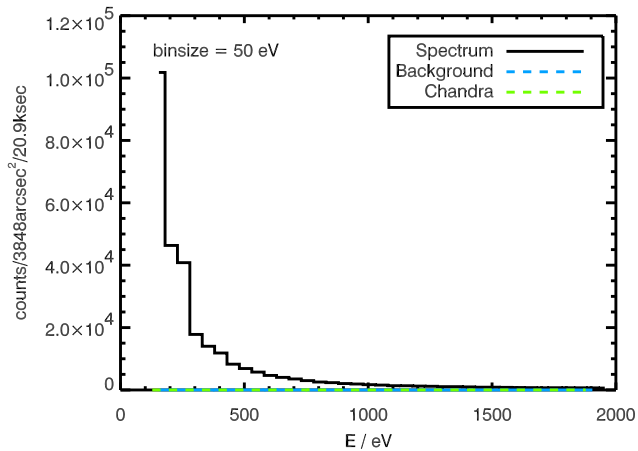}}
\caption{\label{med}Spectrum extracted from the PN observation using the
medium filter (ObsID 0089370601). Clear indications of optical contamination
can be identified.}
\end{figure}

\begin{figure}[!ht]
  \resizebox{\hsize}{!}{\includegraphics{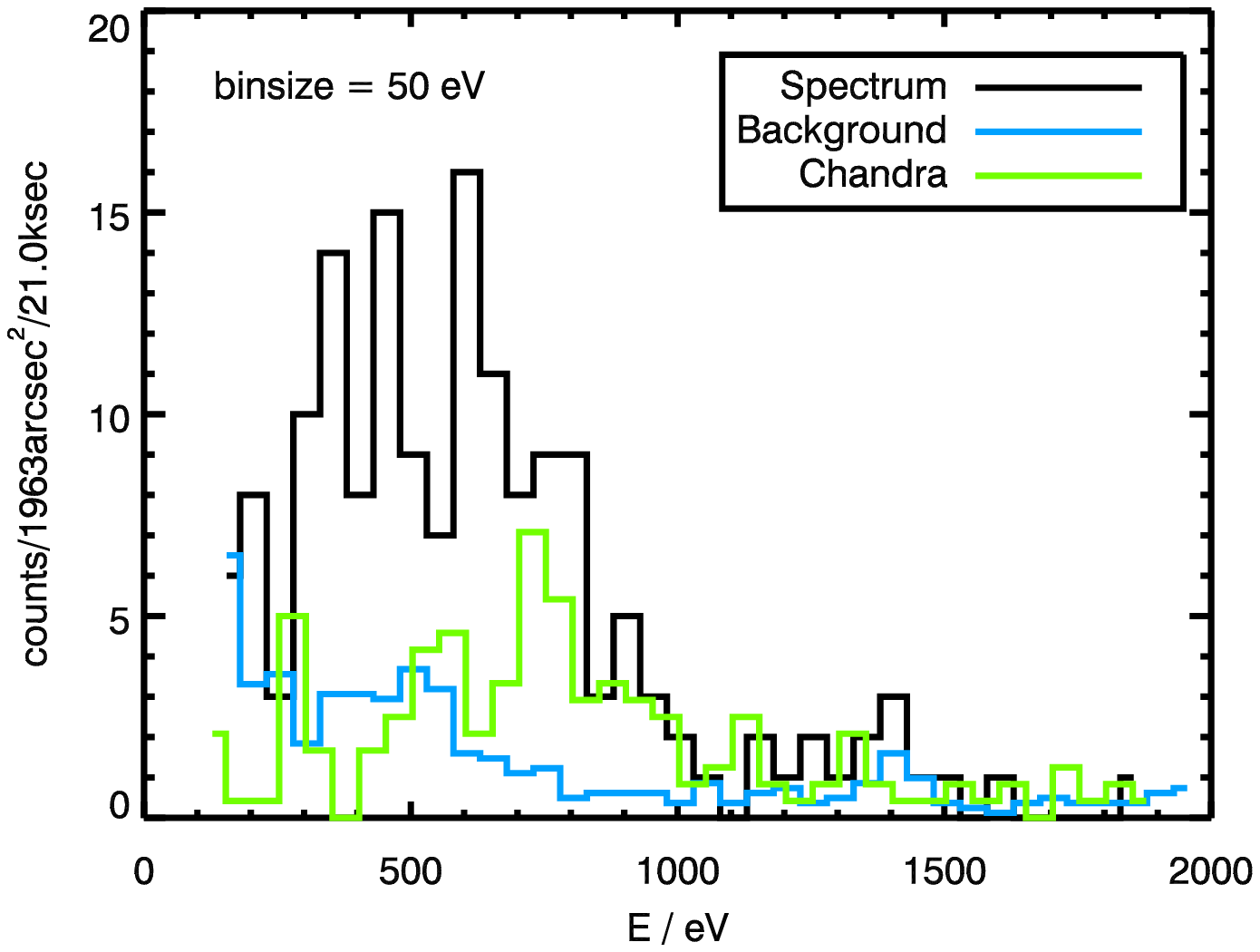}}
  \vspace{-.2cm}\resizebox{\hsize}{!}{\includegraphics{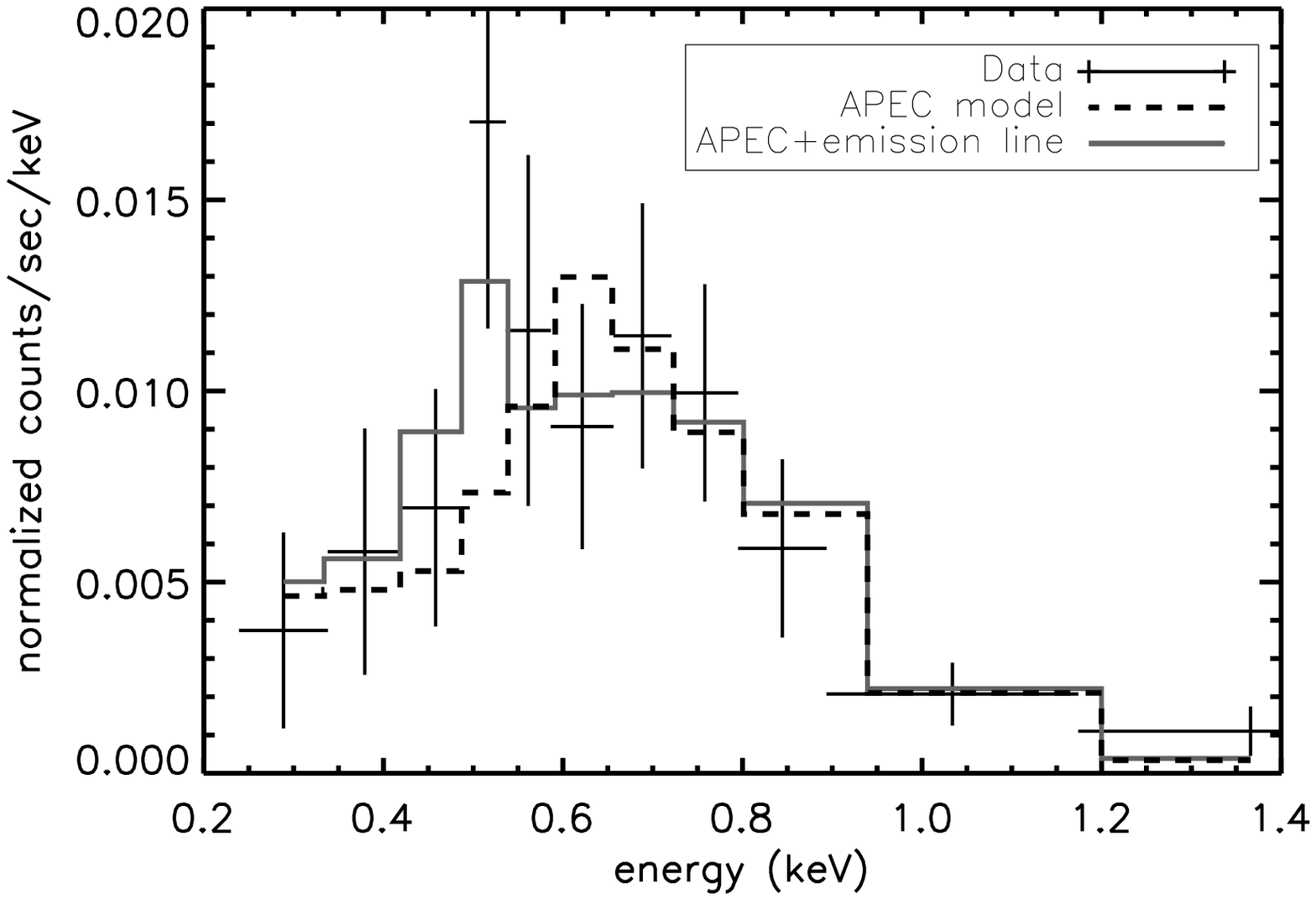}}
\caption{\label{spec}Extracted spectra for the background and source+background
(top). The {\it Chandra} spectrum, scaled to our exposure time and source
extraction area, is overplotted. Bottom: Rebinned spectrum containing at least
15 counts per bin with two models obtained with XSPEC. The best-fit model
consists of an APEC model and an oxygen fluorescent line (see
Table~\ref{model}).}
\end{figure}

Is the signal recorded in the thick filter due to X-rays or also due to
optical contamination? In order to address this issue we analyzed the
EPIC-PN medium filter data in precisely the same fashion as the thick filter
data and extracted a background-subtracted spectrum of the photons attributed to
Saturn (cf. Fig.~\ref{med}). The strong signal increase towards lower energies
is the indicator of the severe contamination due to optical light. In contrast,
carrying out the same procedure with the thick filter data results in a
spectrum looking totally different (cf. Fig.~\ref{spec}). The thick filter
spectrum does not exhibit any increase towards lower energies as expected from a
genuine X-ray spectrum, since the effective areas decrease towards lower
energies. Further, the thick filter spectrum appears very similar to
the recorded {\it Chandra} spectrum, which is overplotted in a light color
after scaling to our exposure time and extraction area. The signal is lower
due to the lower
effective areas of {\it Chandra} mirrors, with the {\it Chandra} spectrum
appearing somewhat shifted towards higher energies. This might be due to some
optical loading in the {\it Chandra} observation, an effect that could not
fully be excluded by \cite{ness_sat}. A rather weak emission feature appears at
$\sim$\,1.3\,keV, but is not significant; interestingly it is also seen in the
{\it Chandra} spectrum.\\

Using XSPEC we carried out spectral modeling similar to \cite{ness_sat}, who
found acceptable spectral fits with a (physically unmotivated) black body model
and a combined MEKAL model plus a fluorescent line of oxygen. MEKAL contains
continuum and line emissivities from collisionally ionized plasma in thermal
equilibrium. This model is supposed to represent the spectrum of the solar corona and the
model parameters are the equilibrium temperature and elemental abundances.
Given the low signal-to-noise of our XMM-Newton data, we can only check
to what extent the XMM-Newton and {\it Chandra} spectra are consistent with each
other. We rebinned the XMM-Newton spectrum to contain at least 15 counts per bin,
necessary to remain outside the Poissonian regime, otherwise a non-standard statistical
treatment is necessary \citep[e.g.,][]{cash79,newi02}. In addition we applied
the Cash statistics provided by XSPEC with the original spectrum and found
consistent results. In order to present a concise goodness-of-fit parameter we here
present our results from $\chi^2$ fits. Our
best-fit black body model yields a temperature of kT$=0.16\,\pm\,0.03$\,keV
($\chi^2_{\rm red}=1.09$ with 9\,dof), consistent with the temperature
found from the {\it Chandra} observation (0.18\,keV). Instead of a MEKAL model we
chose an APEC model to describe an incident solar spectrum. Assuming solar
abundances we find a temperature of kT$=0.29\,\pm\,0.05$\,keV
($\chi^2_{\rm red}=0.79$ with 9\,dof), a little cooler than the
temperature found from the combined MEKAL/fluorescent line model from the
{\it Chandra} spectrum (kT$=0.39\,\pm\,0.08$\,keV). A slightly better fit is obtained
by introducing an oxygen fluorescent line at 527\,eV (modeled as a
narrow emission line, only instrumentally broadened). With this combined
model we obtain an APEC temperature of kT$=0.33\,\pm\,0.08$\,keV,
consistent with the {\it Chandra} results ($\chi^2_{\rm red}=0.41$ with 8\,dof).
The fit results are summarized in Table~\ref{model} and in the last column we
list the model fluxes, integrated in the wavelength interval 0.1--2\,keV.
For an overview of the available X-ray spectra of Saturn we plot
the rebinned spectrum with the APEC model (dashed) and the best-fit model of the
combined APEC/fluorescent line (solid grey) in the bottom panel of Fig.~\ref{spec}.

\begin{table}[!ht]
 \caption{\label{model}Saturn - spectral fits and fluxes}
  \begin{tabular}{lccc}
Model   & kT (keV)$^a$ & $\chi^2_{\rm red}$ / dof$^b$ & flux$^e$\\
{\bf EPIC/PN}&&& {\scriptsize (erg/cm$^2$/s)}\\
 black body &  $0.16\,\pm\,0.03$ & 1.09/9  &  1.66$\cdot10^{-14}$\\
 APEC$^c$   &  $0.29\,\pm\,0.05$ & 0.79/9  &  1.62$\cdot10^{-14}$\\
APEC$^c$+&&&\\
\ \ \ narrow line$^d$& $0.33\,\pm\,0.08$ & 0.41/8 &  1.58$^f\cdot10^{-14}$\\
\hline
\multicolumn{4}{l}{\bf Chandra}\\
black body & 0.18 & 0.7/10 & 0.44$\cdot10^{-14}$\\
MEKAL+ & $0.39\,\pm\,0.08$ & $\downarrow$ & 0.55$\cdot10^{-14}$\\
\ \ \ narrow line& -- & 0.9/9 &  0.13$\cdot10^{-14}$\\
{\bf ROSAT} & -- & --& 1.9$\cdot10^{-14}$\\
\hline
  \end{tabular}
\\
$^a$90\% errors \hspace{2cm} $^b$degrees of freedom\\
$^c$solar abundances \hspace{.6cm} $^d$at 527\,eV, delta profile
\hspace{.3cm} $^e$0.1--2\,keV\\
$^f$1.36$\cdot10^{-14}$ (APEC)\,+\,0.22$\cdot10^{-14}$ (fluorescent line)
\end{table}

While none of the considered spectral models may be physically correct, they
allow a reasonable accurate estimate of the X-ray flux recorded by XMM-Newton.
An inspection of Table~\ref{model} shows, that an apparent X-ray flux between
$1.58-1.66\cdot10^{-14}$\,erg\,cm$^{-2}$\,s$^{-1}$ was recorded by XMM-Newton.
This compares well with the ROSAT flux of
$1.9\cdot10^{-14}$\,erg\,cm$^{-2}$\,s$^{-1}$ (in the soft ROSAT band
0.1--0.55\,keV) reported by \cite{ness00}, but is above the level measured by
{\it Chandra} of $0.68\cdot10^{-14}$\,erg\,cm$^{-2}$\,s$^{-1}$. Thus an
interpretation of the signal recorded by XMM-Newton in the thick filter as
genuine X-ray emission yields flux values consistent with earlier observations
of X-ray emission from Saturn as well as an X-ray spectrum consistent with that
recorded by {\it Chandra}. We therefore conclude that indeed true X-ray emission
from Saturn has been recorded by XMM-Newton.\\

\section{Discussion and Conclusions}

We analyzed XMM-Newton observations of Saturn. Because of Saturn's
visual magnitude of $m_V = 0.9$ at the time of our observations, the
data taken with the thin and the medium filters are severely contaminated by
optical light. However, the data taken with the thick filter are
exclusively X-ray photons originating from Saturn. Thus, X-ray emission from
Saturn has been established to be significantly weaker than
for Jupiter; the reported flux levels range from
$0.68\cdot10^{-14}$\,erg\,cm$^{-2}$\,s$^{-1}$ from {\it Chandra} to
$1.9\cdot10^{-14}$\,erg\,cm$^{-2}$\,s$^{-1}$ in a marginal ROSAT detection
\citep{ness00}. At least between the {\it Chandra} and XMM-Newton observations
substantial variablity seems to have taken place, a fact hardly surprising
for almost any X-ray source. The spectral models found to be consistent
with the XMM-Newton data are also consistent (to within the errors) with the
results from the {\it Chandra} observation. Any possibly remaining problems with
optical loading do not seem to significantly affect the results.
Further insights into Saturn's X-ray production process require substantially
deeper pointings than presently available.

\begin{acknowledgements}
This work is based on observations obtained with XMM-Newton, an ESA science
mission with instruments and contributions directly funded by ESA Member
States and the USA (NASA). We also thank Pedro Rodriguez from the XMM helpdesk
for his assistance.\\
J.-U.N. and J.R. acknowledge support from DLR under 50OR0105.

\end{acknowledgements}

\bibliographystyle{aa}
\bibliography{astron,jn,sat,jhmm}

\end{document}